\documentstyle[12pt,fleqn]{article}

\def\soc{{\rm C}_{60}}
\def\rug{{\rm C}_{70}}
\def\beeq{\begin{equation}}
\def\eneq{\end{equation}}
\def\beeqa{\begin{eqnarray}}
\def\eneqa{\end{eqnarray}}

\addtocounter{section}{-1}
\begin{document}

\begin{center}

{\large {\bf Polaron excitations in doped $\soc$:\\
Effects of disorders}\\
 }

\vspace{1cm}

{\rm Kikuo Harigaya$^*$}\\

\vspace{1cm}

{\sl Department of Physics, University of Sheffield,\\
Sheffield S3 7RH, United Kingdom}\\
and\\
{\sl Fundamental Physics Section, Physical Science Division,\\
Electrotechnical Laboratory,\\
Umezono 1-1-4, Tsukuba, Ibaraki 305, Japan$^{**}$}

\vspace{1cm}

(Received March 12, 1993)
\end{center}

\vspace{1cm}

\noindent
{\bf ABSTRACT}\\
Effects on C$_{\rm 60}$ by thermal fluctuations of phonons,
misalignment of C$_{\rm 60}$ molecules in a crystal, and other
intercalated impurities (remaining C$_{\rm 70}$, oxygens, and
so on) are simulated by disorder potentials.  The
Su-Schrieffer-Heeger--type electron-phonon model for doped
C$_{\rm 60}$ is solved with gaussian bond disorders and also
with site disorders.  Sample average is performed over
sufficient number of disorder configurations.  The distributions
of bond lengths and electron densities are shown as functions
of the disorder strength and the additional electron number.
Stability of polaron excitations as well as dimerization patterns
is studied.  Polarons and dimerizations in lightly doped cases
(C$_{\rm 60}^{-1,-2}$) are relatively stable against disorders,
indicated by peak structures in distribution functions.
In more heavily doped cases, the several peaks merge into a
single peak, showing the breakdown of polaron structures as
well as the decrease of the dimerization strength.  Possibility
of the observation of polaronic lattice distortions and electron
structures in doped C$_{\rm 60}$ is discussed.

{}~


\pagebreak


\section{INTRODUCTION}

Recently, the fullerenes C$_N$ which have the hollow cage
structures of carbons have been intensively investigated.
There are several experimental indications that the doped
fullerenes show polaronic properties due to the Jahn-Teller
distortion, for example: (1) The electron spin resonance (ESR)
study [1] on the radical anion of C$_{60}$ has revealed the
small $g$-factor, $g=1.9991$, and this is associated with
the residual orbital angular momentum due to the Jahn-Teller
distortion. (2) Photoemission studies [2] of C$_{60}$ and
C$_{70}$ doped with alkali metals have shown peak structures,
which cannot be described by a simple band-filling picture.
(3) When poly(3-alkylthiophene) is doped with $\soc$ [3],
interband absorption of the polymer is remarkably suppressed
and the new absorption peak evolves in the low energy range.
The Jahn-Teller splitting of LUMO in C$_{60}^-$ state and/or
the Coulomb attraction of positively charged polaron to
C$_{60}^-$ might occur. (4) The luminescence of neutral $\soc$
has been measured [4].  There are two peaks around 1.5 and
1.7eV below the gap energy 1.9eV, interpreted by the effect
of the polaron exciton. In addition, the experiments on the
dynamics of photoexcited states have shown the interesting
roles of polarons [5].

Several authors [6,7] have proposed an interacting
electron-phonon model in order to describe the polarons
in doped $\soc$.  The Su-Schrieffer-Heeger (SSH) model
of conjugated polymers [8] has been extended to fullerenes
in these works [6,7].  The $\pi$-electrons hop between nearest
neighbor sites.  The hopping integral depends on the change
of the bond length linearly.  The bond modelled by the classical
harmonic spring is the contribution from the $\sigma$ bonding.
In the previous paper [9], we have calculated
lattice distortion and electronic structures of the molecules,
where one to six electrons are added, or one to ten electrons
are removed.  When $\soc$ is doped with one or two electrons
(or holes) (the lightly doped case), the additional charges
accumulate at twenty carbon atoms along almost an equatorial
line of the molecule.  The dimerization becomes the weakest
along the same line.  Two energy levels, the occupied state
and the empty state, intrude largely in the gap.  These are
the polaron effects.  The changes of the electronic structures
of the molecules with more charges (the heavily doped case)
have been reported in Ref. 9.  However, the complex changes of
lattice geometries and electron density distributions have not
shown yet.  Section III of this paper will be devoted
to this purpose.

In the study [9], the molecule has been assumed as isolated
and the calculations have been done within the adiabatic
approximation.  However, it has been discussed [10] that the
width of the zero point motion in conjugated polymers and
fullerene tubules is of the order of 0.01\AA.  The same order
of magnitude would be expected in $\soc$ [11].  This is also
of the same order as the difference between the short and
bond lengths: 0.05\AA{\ } [12].  Therefore, the polaronic
distortion described in the adiabatic approximation might
change its structures by thermal fluctuations.  The thermal
fluctuation effects can be simulated by introducing the bond
disorder potentials.  The doped SSH system with gaussian
bond disorders will be studied in Sec. IV.

It is known that $\soc$ molecules contain a small amount of
$\rug$ as impurities [4].  Sometimes the $\soc$ films and
solids are contaminated with oxygens [4,13].  There would
remain misalignment of molecules in $\soc$ solids.  These
effects would be good origins of additional potentials acting
on $\pi$-electrons at the carbon sites.  They can be modelled
by random site disorders.  Site disorder effects will be
investigated in Sec. V.

Sample average is performed over sufficient number of independent
disorder configurations.  The distribution functions of bond
lengths and electron densities are calculated with changing
the disorder strength and the additional electron number.
We mainly consider stability of polaron excitations as well
as dimerization patterns.   We find that polarons and
dimerizations in lightly doped cases are rather stable against
disorders.  This property is common to bond and site disorder
effects.  In more heavily doped cases, the several peaks in
distribution functions merge into a single peak.  This
indicates that polaron structures are broken while the
dimerization strength decreases, owing to doped charges
and disorders.

This paper is organized as follows.  We explain the model in the next
section.  In Sec. III, the lattice and electronic patterns of
electron-doped $\soc$ are extensively reported.  In Sec. IV,
bond disorder effects are studied.   Sec. V is devoted to site
disorder effects.  We close this paper with summary and discussion
in Sec. VI.

\section{MODEL}

The extended SSH hamiltonian for the fullerene $\soc$ [9],
\beeq
H_{\rm SSH} = \sum_{\langle i,j \rangle, \sigma} ( - t_0 + \alpha y_{i,j} )
( c_{i,\sigma}^\dagger c_{j,\sigma} + {\rm h.c.} )
+ \frac{K}{2} \sum_{\langle i,j \rangle} y_{i,j}^2,\\
\eneq
is studied with gaussian bond disorders,
\beeq
H_{\rm bond} = \alpha \sum_{\langle i,j \rangle, \sigma} \delta y_{i,j}
( c_{i,\sigma}^\dagger c_{j,\sigma} + {\rm h.c.} ),
\eneq
as well as with gaussian site disorders,
\beeq
H_{\rm site} = \sum_{i,\sigma} U_i c_{i,\sigma}^\dagger c_{i,\sigma}.
\eneq
In $H_{\rm SSH}$, $c_{i,\sigma}$ is an annihilation operator of a
$\pi$-electron; the quantity $t_0$ is the hopping integral of the ideal
undimerized system; $\alpha$ is the electron-phonon coupling; $y_{i,j}$
indicates the bond variable which measures the length change of the bond
between the $i$- and $j$-th sites from that of the undimerized system;
the sum is taken over nearest neighbor pairs $\langle i j \rangle$; the
second term is the elastic energy of the lattice; and the quantity $K$
is the spring constant.  The part $H_{\rm bond}$ is the disorder
potential due to the gaussian modulation of transfer integrals.  The
disorder strength is measured by the standard deviation $y_s$ of the
bond variable modulations $\delta y_{i,j}$.  The mean value of $\delta
y_{i,j}$ is assumed to be zero.  The term $H_{\rm site}$ is the site
disorder potential.  The quantity $U_i$ is the strength of the disorder
at the $i$th site, with the standard deviation $U_s$.  The model is
solved with the assumption of the adiabatic approximation and by an
iteration method used in [9].

\section{POLARONS IN C$_{\rm 60}$}

In this section, we present detailed discussion on the
polarons of the electron doped $\soc$.  Further data of the
lattice and electron density structures are reported.
In Ref. 9, only the electronic energy levels and averaged
dimerization strengths have been discussed for heavily doped $\soc$.

In Fig. 1, the unfolded figure of the truncated icosahedron
is shown.  When we make the paper model of $\soc$, we cut
the figure along the outer edges of white hexagons.  After
folding edges between neighboring hexagons and combining
several edges, we obtain a closed structure of the molecule.
The shadowed hexagons in Fig. 1 become pentagons.
The symbols, A-L, specify different pentagons.

The model Eq. (1) is numerically solved for the parameters:
$t_0 = 2.1$eV, $\alpha = 6.0$eV/\AA, and $K = 52.5$eV/\AA$^2$.
These give the characteristic scales for the neutral $\soc$:
the total $\pi$-band width $6t_0 = 12.6$eV, the energy gap
1.904eV, and the difference of the bond length between the
short and long bonds 0.4557\AA.  These values are typical.
They are slightly different from those in Ref. 9, but the
qualitative features of solutions are not affected by the
slight modifications.  Quantitative differences are small, too.
The total electron number is changed within $N \leq N_{\rm el}
\leq N+6$, $N = 60$ being the number of sites.

In Fig. 2, the magnitudes of bond variables are shown in upper
figures by changing the electron number.  The bond is represented
by a bold line when it is shorter and $y_{i,j} < 0$.  The bond
is represented by a dashed line when it is longer and
$y_{i,j} > 0$.  The thickness of the line indicates
$|y_{i,j}|$.  The valence number (the negative of the
additional electron number) of the doped molecule is
accompanied with each figure.  The lower figures show the
additional electron density.  The area in each circle is
proportional to the absolute value.

In $\soc^{-1,-2}$, the most (about 70-80 percent) of the
additional charges accumulate at twenty carbons along an
equatorial line of the molecule.  At the same time, the
difference between the lengths of the short and long bonds
becomes smallest at sites along the same equatorial line.
This means that the dimerization strength is weakest.  Two
nondegenerate energy levels intrude largely in the energy
gap as shown in Fig. 2(a) of Ref. 9.  These lattice and
electronic structures are the same as those of polarons in
conjugated polymers, so we named the changes as polaron
excitations.  A fivefold axis penetrates between the centers
of the pentagons, A and H.

When $\soc$ is doped with three electrons, the symmetry is highly
reduced.  There is only an inversion symmetry.  Only two sites have
the same electron density.  Only two bonds have the same length.
As shown in Fig. 2, additional electron densities have large values
at the twenty carbons along the equatorial line as well as
at the other sites near the pentagons, A and H.  The energy levels
are shown in Ref. 9.  The 31th wavefunction has large amplitudes
at sites along the equatorial line, while the 32th one has larger
amplitudes at the other sites.  The distribution of the electron
density reflects this property.

When the molecule is further doped and the additional electron
number is four, the dimerization pattern changes qualitatively.
The symmetry becomes higher and there is a threefold axis
which penetrates the center of the hexagon surrounded by the
pentagons, B, C, and L.  The molecule doped with five electrons
has the similar bond alternation pattern and the same symmetry
(the figures are not shown for the simplicity).
Finally, in the molecule doped with six electrons, the dimerization
almost disappears and is rather negligible.  The icosahedral symmetry
recovers again.

\section{BOND-DISORDER EFFECTS}

The SSH model $H_{\rm SSH}$ is solved with the bond disorders
$H_{\rm bond}$.  The additional electron number,
$n \equiv N_{\rm el} - N$, is changed within $0 \leq n \leq 6$.
The most realistic origin for bond disorders is the thermal
fluctuation of phonons.  So, we shall change the strength of
the disorder in the range comparable with that of the amplitude
of thermal fluctuations.  It has been discussed [10] that the width
of the zero point motion of phonons is 0.03-0.05\AA{\ } in
conjugated polymers and is of the same order in fullerenes.
So, it is reasonable to assume that the maximum value
of $y_s$ is of the similar magnitude.  We take $y_s = 0.01,
0.03,$ and $0.05$\AA.

A fairly large number of mutually independent samples of
disorders are generated, and the model, Eqs. (1) and (2),
is solved for each sample.  The bond variable and electron
density are counted in order to draw histograms of distributions.
The sample number 5000 yields good convergence.

Figure 3 shows the distributions of bond variables $D(y)$.
The thick, thin, and dashed lines are for $y_s = 0.01, 0.03$,
and $0.05$\AA, respectively.  The ordinate is normalized so
that the area between the curve and the abscissa is unity.
Figure 3(a) for the neutral $\soc$ shows the two peak
structure, related with presence of the dimerization: the
positive $y_{i,j}$ corresponds to the longer bond, while
the negative one to the short bond.  The magnitude of the
zero point motion would be of the order 0.01\AA.
For example, the treatment of the $H_g$-type phonon within
the framework of the SSH model by Friedman and Harigaya has
resulted that the magnitude is about 0.02-0.03\AA{\ } as the
adiabatic energy curve in Ref. 11  indicates.  The curve of
$y_s = 0.01$\AA{\ } has two peaks which are clearly separated.
The curve of $y_s = 0.03$\AA{\ } still shows distinct peaks.
Therefore, the dimerization survives thermal fluctuations
in the neutral molecule.  Actually, the nuclear magnetic
resonance (NMR) [12] shows the existence of the dimerization.  When doped
further up to $\soc^{-1,-2}$, the dimerization seems to remain
against disorders:  the two peaks can be identified.   However,
a new peak emerges between the two peaks for $y_s = 0.01$\AA{\ }
when doped with two electrons.  This peak corresponds to the
small bond variables which have been located along the
equatorial line in the impurity free case.  Thus,
{\sl the polaronic distortion seems to persist}, too.

Figure 4 shows the distributions of the electron density per site
$D(\rho)$.  In $\rho$, the electron density of the impurity-free
half-filled system is subtracted.  The notations of the lines
are the same as in Fig. 3.  Figure 4(a) shows the nearly unform
electron density.  When doped with one or two electrons,
a shoulder develops at the positive-$\rho$ side of the curve.
This is owing to the accumulation of the extra charge in the
limited carbon sites of the molecule as found in Fig. 2.
The dimerization begins to be broken in the same portion.
And also, the central peak around $\rho = 0$ still remains,
due to the smaller changes of the electron density at sites
of the pentagons, A and H, of Fig. 2.  Thus, {\sl the polaronic
charge distribution persists in the presence of bond disorders}.

Next, we discuss heavily doped cases with $n \geq 3$.
In the distribution function of the bond variables, the three peaks
merge into a single peak centered around $y = 0$\AA.  This is
owing to the dimerization, the strength of which became very
smaller.  The polaronic distortion becomes smaller also, as indicated
by the averaged dimerization $\langle | y_{i,j} | \rangle $
presented in Table III of Ref. 9.  The electron distribution
functions shown in Figs. 4(d) and (e) have a largest peak
centering the value $n / N$ which is the
result of the uniform doping.   This is also related with the breakdown
of the bond alternation pattern.

\section{SITE-DISORDER EFFECTS}

The model Eqs. (1) and (3) is solved for each sample of site
disorder potentials.  Taking the number of samples up to 5000
yields nice convergence of distribution functions.  We assume three
values for disorder strength: $U_s = 0.5, 1.0,$ and $2.0$eV.
The calculation of Madelung potential in the solid $\soc$ [14]
has yielded the variation of the potential on the surface
of $\soc$ within 0.5eV.  Therefore, the misalignment of $\soc$
in the solid gives rise to the similar strength of site
disorders.  The other intercalated impurities (remaining
C$_{\rm 70}$ [4], oxygens [4,13], dopants [2], and so on)
might yield site disorder potentials of the order of 1eV.
These are the realistic origins of site disorders.
The additional electron number $n$ is changed up to six.

Fig. 5 shows the distribution functions of bond variables $D(y)$.
The thick, thin, and dashed lines are for $U_s = 0.5, 1.0$,
and $2.0$eV, respectively.  While the molecule is weakly doped
($n = 1,2$), the dimerization tends to survive disorder potentials.
This is easily seen by thick and thin lines in Figs. 5(b) and (c).
The dashed lines have a single peak.  This is due to the strong
disorder potential comparable to the size of the energy gap of the undoped
molecule; the dimerization becomes undisernible, when energies of the
occupied and unoccupied electronic states are closer.  The actual
site potentials would not so strong as that of the dashed line,
in view of the screening effects due to the $\pi$-electrons spread
over the surface of the molecule.  Therefore, {\sl the dimerization
persists strongly when site disorders are present}.

When the doping proceeds further ($3 \leq n \leq 6$), the two major
peaks join into a single peak in $D(y)$.  This shows that the dimerization
is easy to disappear due to the disorder potentials as well as
the densely accumulated extra charges.

Figure 6 shows charge density distributions $D(\rho)$.
We show results only for $n = 0, 3,$ and 6, because the qualitative
features are the same for all $n$.  The notation of the lines is the same
as in Fig. 5.  The charge density is directly modulated by the
site disorders.  So, each curve has the shape near the gaussian
distribution.  The value of $\rho$ at the peak is close to $n/N$.

\section{SUMMARY AND DISCUSSION}

Effects on C$_{\rm 60}$ by thermal fluctuations of phonons
have been simulated by bond disorder potentials.  Next,
misalignment of C$_{\rm 60}$ molecules in a crystal, and
other intercalated impurities (remaining C$_{\rm 70}$,
oxygens, dopants, and so on) have been studied with site
disorder potentials.  The extended SSH model for doped
C$_{\rm 60}$ has been solved with the assumption of the
adiabatic approximation.  The distributions of bond lengths
and electron densities, $D(y)$ and $D(\rho)$, have been
shown as functions of the disorder strength and the
additional electron number.  Stability of polaron excitations
as well as dimerization patterns have been considered.

Main conclusions are common to bond and site disorder effects.
Polarons and dimerizations in lightly doped cases
(C$_{\rm 60}^{-1,-2}$) are relatively stable against disorders.
This property has been indicated by peak structures in
distribution functions.  In more heavily doped cases, the
several peaks merge into a single peak, showing the
breakdown of polaron structures as well as the decrease
of the dimerization strength.

However, there exist qualitative differences between bond and site
disorder effects.  In the bond disorder problem, the bond length is
affected directly by the disorder potentials, but the charge density
is modulated indirectly.  In the site disorder problem, the charge
density is modulated directly by the disorder potentials.  Therefore,
the distribution function of charge density shows the apparent peak
structure related with the dimerization and polaronic distribution in
the bond disorder problem, but it has only one peak in the site
disorder problem.

Then, how is our finding related with experiments?  The NMR
investigation [12] gives the evidence that there are two bond lengths
in $\soc$.  The single molecule will be always in the presence
of some kinds of external potentials.  These potentials could be
effectively regarded as disorders.  The presence of the two bond
lengths in actual molecules agrees with our result that the
dimerization is relatively stable against disorders.  The ESR
study [1] of $\soc$ monoanion in the solvent shows the reduced
$g$-factor.  This is interpreted as the result of the Jahn-Teller
distortion, in other words, the polaronic distortion.  The molecules
in the solvent would be affected by the strong site disorders as
well as bond disorders.  Nevertheless, the effect related with
polarons is observed.  This is again in accord with our
conclusion that polarons are stable in lightly doped $\soc$ in
disordered external potentials.

The electrochemical experiment [15] can produce $\soc$ anions doped with
up to six electrons.  The molecule can be doped with six electrons
in the solid also.  The photoemission experiments [2] show that the
maximumly doped $\soc$ solid is an insulator.  This accords with
the present calculation.  However, it is not certain whether the
dimerizations still remain or not in heavily doped $\soc$.
In view of the fact that dimerizations are very small in the heavily
doped $\soc$ and the width of zero point motion of phonons is the
order of 0.01\AA{\ } [11], it is certainly possible that the dimerization
would not be observed in actual samples.

{}~~~~~~

\noindent
{\bf ACKNOWLEDGEMENTS}\\
Fruitful discussion with Prof. G. A. Gehring, Dr. M. Fujita,
and Dr. Y. Asai is acknowledged.  Useful correspondences with
Prof. B. Friedman and Dr. S. Abe are also acknowledged.
The author is grateful to Dr. M. Fujita
for providing him with Figs. 1 and 2 of this
paper.  Numerical calculations have been performed on FACOM
M-1800/30 of the Research Information Processing System,
Agency of Industrial Science and Technology, Japan.

\pagebreak

\begin{flushleft}
{\bf REFERENCES}
\end{flushleft}

\noindent
$*$ electronic mail address: harigaya@etl.go.jp.\\
$**$ permanent address.\\
$[1]$ T. Kato, T. Kodama, M. Oyama, S. O\-ka\-za\-ki, T. Shida, T. Nakagawa,
Y. Matsui, S. Suzuki, H. Shiromaru, K. Yamauchi, and Y. Achiba,
Chem. Phys. Lett. {\bf 180}, 446 (1991).\\
$[2]$ T. Takahashi, S. Suzuki, T. Morikawa,  H. Katayama-Yoshida,
S. Hasegawa, H. Inokuchi, K. Seki, K. Kikuchi, S. Suzuki, K. Ikemoto,
and Y. Achiba, Phys. Rev. Lett. {\bf 68}, 1232 (1992);
C. T. Chen, L H. Tjeng, P. Rudolf, G. Meigs, L. E. Rowe, J. Chen, J. P.
McCauley Jr., A. B. Smith III, A. R. McGhie, W. J. Romanow,
and E. W. Plummer, Nature {\bf 352}, 603 (1991).\\
$[3]$ S. Morita, A. A. Zakhidov, and K. Yoshino, Solid State Commun. {\bf 82},
249 (1992); S. Morita, A. A. Zakhidov, T. Kawai, H. Araki, and K. Yoshino,
Jpn. J. Appl. Phys. {\bf 31}, L890 (1992).\\
$[4]$ M. Matus, H. Kuzmany, and E. Sohmen, Phys. Rev. Lett. {\bf 68},
2822 (1992).\\
$[5]$ P. A. Lane, L. S. Swanson, Q. X. Ni, J. Shinar, J. P. Engel,
T. J. Barton, and L. Jones, Phys. Rev. Lett. {\bf 68}, 887 (1992).\\
$[6]$ F. C. Zhang, M. Ogata, and T. M. Rice, Phys. Rev. Lett.
{\bf 67}, 3452 (1991).\\
$[7]$ K. Harigaya, J. Phys. Soc. Jpn. {\bf 60}, 4001 (1991);
B. Friedman, Phys. Rev. B {\bf 45}, 1454 (1992).\\
$[8]$ W. P. Su, J. R. Schrieffer, and A. J. Heeger, Phys. Rev. B {\bf 22},
2099 (1980).\\
$[9]$ K. Harigaya, Phys. Rev. B {\bf 45}, 13676 (1992).\\
$[10]$ McKenzie and Wilkins, Phys. Rev. Lett. {\bf 69}, 1085 (1992).\\
$[11]$ B. Friedman and K. Harigaya, Phys. Rev. B {\bf 47},
(1993) (February issue; in press).\\
$[12]$ C. S. Yannoni, P. P. Bernier, D. S. Bethune,
G. Meijer and J. R. Salem, J. Am. Chem. Soc. {\bf 113}, 3190 (1991).\\
$[13]$ T. Arai, Y. Murakami, H. Suematsu, K. Kikuchi, Y. Achiba,
and I. Ikemoto, Solid State Commun. {\bf 84}, 827 (1992).\\
$[14]$ K. Harigaya, (preprint).\\
$[15]$ Q. Xie, E. P\'{e}rez-Cordero, and L. Echegoyen,
J. Am. Chem. Soc. {\bf 114}, 3978 (1992).\\

\pagebreak

\begin{flushleft}
{\bf Figure Captions}
\end{flushleft}

\noindent
FIG. 1.  Unfolded pattern of the paper model of $\soc$.
See the text for the notations.

{}~

\noindent
FIG. 2.  Bond variables and excess electron densities shown on the
unfolded pattern of doped $\soc$ without disorders.  The upper figures
show the bond variables, while lower figures display excess electron
densities.  The number at the top is $N - N_{\rm el}$.   Notations
are explained in the text.

{}~

\noindent
FIG. 3.  Distribution function $D(y)$ of bond variables $y$ of
the doped $\soc$ in the presence of bond disorders.
The ordinate is normalized so that the area between the curve
and the abscissa is unity.  The thick, thin, and dashed lines
are for $y_s = 0.01, 0.03$, and $0.05$\AA, respectively.

{}~

\noindent
FIG. 4.  Distribution function $D(\rho)$ of the excess electron density
of the doped $\soc$ in the presence of bond disorders.
The ordinate is normalized so that the area between the curve
and the abscissa is unity.  The thick, thin, and dashed lines
are for $y_s = 0.01, 0.03$, and $0.05$\AA, respectively.

{}~

\noindent
FIG. 5.  Distribution function $D(y)$ of bond variables $y$ of
the doped $\soc$ in the presence of site disorders.
The ordinate is normalized so that the area between the curve
and the abscissa is unity.  The thick, thin, and dashed lines are for
$U_s = 0.5, 1.0$, and $2.0$eV, respectively.

{}~

\noindent
FIG. 6.  Distribution function $D(\rho)$ of the excess electron density
of the doped $\soc$ in the presence of site disorders.
The ordinate is normalized so that the area between the curve
and the abscissa is unity.  The thick, thin, and dashed lines are for
$U_s = 0.5, 1.0$, and $2.0$eV, respectively.


\end{document}